\documentclass[conference, compsoc]{IEEEtran}
\IEEEoverridecommandlockouts
\usepackage{subfigure}
\usepackage{caption}
\usepackage{hyperref}

\usepackage{amsmath}
\usepackage{booktabs}
\usepackage{graphicx}
\usepackage{multirow}
\usepackage{courier}
\usepackage{multirow}
\usepackage{acronym}
\usepackage{color}

\begin{document}

\title{Internet-based Adaptive Distributed Simulation\\ of Mobile Ad-hoc Networks\footnotemark}

\author{
\IEEEauthorblockN{Gabriele D'Angelo, Stefano Ferretti}
\IEEEauthorblockA{Department of Computer Science and Engineering (DISI)\\University of Bologna, Italy\\
\{g.dangelo, s.ferretti\}@unibo.it}
\and
\IEEEauthorblockN{Gary S. H. Tan}
\IEEEauthorblockA{Department of Computer Science\\School of Computing\\National University of Singapore, Singapore\\
dcstansh@nus.edu.sg}
}

\maketitle

\footnotetext{
The publisher version of this paper is available at \url{https://doi.org/10.1109/WSC40007.2019.9004796}.
\textbf{{\color{red} This is the pre-peer reviewed version of the article: ``Gabriele D'Angelo, Stefano Ferretti, Gary S. H. Tan. Internet-based Adaptive Distributed Simulation of Mobile Ad-hoc Networks. Proceedings of the Proceedings of the 2019 Winter Simulation Conference (WSC 2019).''.}}}

\begin{abstract}
In this paper we focus on Internet-based simulation, a form of distributed simulation in which a set of execution units that are physically located around the globe work together to run a simulation model. This setup is very challenging because of the latency/variability of communications. Thus, clever mechanisms must be adopted in the distributed simulation, such as the adaptive partitioning of the simulated model and load balancing strategies among execution units. We simulate a wireless model over a real Internet-based distributed simulation setup, and evaluate the scalability of the simulator with and without the use of adaptive strategies for both communication overhead reduction and load-balancing enhancement. The results confirm the viability of our approach to build Internet-based simulations.\\
\end{abstract}

\section{INTRODUCTION}
\label{sec:intro}

Distributed simulation, and in particular Internet-based simulation, has long been neglected for several reasons. A primary motivation is the network latency that is typical of Internet communications. While the introduction of novel communication technologies has reduced the average Internet delays, latencies for the transfer of bulk data do have significant costs. As a confirmation, take the recent reconstruction of a black hole image that has been produced via analysis of 4 petabytes~\cite{blackhole}. The choice to transmit all the collected data from multiple sources around the world was to ship the (costly) hard drives, rather than sending data over the Internet as it was faster and safer to transmit the hardware, rather than the data. Reliability is another reason. The ability to cope with node and network failures is a main research issue in distributed systems. While several solutions have been identified, simulationists still seem to prefer not to use a distributed simulation environment~\cite{gda-dsrt-2016}. A third issue is related to the security aspects of Internet-based simulation. Confidentiality and anonymity play an important role in certain simulation contexts, e.g.~highly classified warfare simulation~\cite{gda-dsrt-2018a}. In a distributed simulation, how can we make sure that the data exchanged among the hosts cannot be intercepted or tampered with? The trivial solution to this problem is to centralize all simulation components in a single location.

However, the possibility of creating real distributed simulations remains an important goal to pursue~\cite{FUJ00}. It is not common to transfer bunch of petabytes at a time to perform a simulation run. Conversely, in a simulation, we do often need a high amount of memory and computational resources, that make the use of multiple execution units a very interesting possibility. Quite often, these resources might be distributed, when it is not possible to create a viable simulation setup using the hardware available in a single computation center. Distributed simulation is a technique that can be useful to build scalable simulations by aggregating a pool of distributed resources. Large simulations could be realized by federating components that are owned or controlled by different organizations who may not be willing to run their code out of their premises. Moreover, the possibility to take advantage of (multi-vendor) cloud services, which allow a dynamic and rapid re-configuration of the computation capacity, by adding and removing working nodes on demand, is fundamental for building simulation models with a certain degree of complexity.

The Parallel And Distributed Simulation (PADS) community usually focuses on the use of parallel computation units. When focusing on the ``distribution'', usually Logical Processors (LPs), i.e.~the simulation model components in the PADS jargon, are placed in a low-latency cluster or a local area networks, rather than in wide area networks. We are aware of just few works that provide an experimental evaluation of a PADS over the Internet~\cite{746044-1998,Lu:2000:SLD:361026.361034,pads}.

We do believe that more emphasis should be put on Internet-based simulations in order to study the existing distributed simulation mechanisms when in presence of high-latency, low bandwidth and high jitter networks. Moreover to fully support viable solutions, sophisticated additional mechanisms should be employed. For example, partitioning and adaptive migration strategies could reduce the communication needs among distributed LPs, as well as balance the computational workload among different LPs~\cite{gda-simpat-2017}. This is of primary importance, due to the infeasibility of statically partitioning certain simulation models, and the possible variability of both network and execution units performance in Internet-based execution environments.

In this paper we present results from an experimental evaluation of an Internet-based distributed simulation built on top of the simulation middleware~\cite{gda-simpat-2014} and the adaptive reallocation framework~\cite{gda-simpat-2017} that we proposed in our previous research works. For the analysis, we employed a time-stepped~\cite{FUJ00}, ad hoc wireless network model, typical of an Internet of Things (IoT) scenario. Nodes move in a simulated area through a random waypoint mobility model and communicate via a local broadcast, using an ad hoc wireless communication technology that allows reaching nodes within a specific transmission range. To evaluate the performance of the Internet-based distributed simulator, we preferred to rely on a real-world testbed instead of building a synthetic environment with simulated network performance and background load. The testbed is composed of 3 virtual machines executed in different continents (i.e.~Europe, Asia and America). Furthermore, we have chosen virtual machines with heterogeneous specifications (e.g.~number of CPU cores) to test the behavior of the simulator under these circumstances. Clearly, this execution setup is quite challenging with respect to the LAN environments that we employed in past studies~\cite{gda-simpat-2017}. In other words, the main contribution of this paper is the evaluation of the distributed simulation and adaptive reallocation strategies in a Internet-based setup.

The remainder of this paper is organized as follows. Section~\ref{sec:back} describes the background related to the paper subject. Section~\ref{sec:related} outlines the state of the art and related work in the field.  Section~\ref{sec:partitioning} discusses the main schemes used to perform simulation model partitioning. The methodology employed in the performance evaluation and related results are described in Section~\ref{sec:perf}. Finally, Section~\ref{sec:conc} provides some concluding remarks.

\section{BACKGROUND}\label{sec:back}

In this section, we introduce the main concepts of simulation, focusing on discrete event simulation, Internet-based and distributed simulation techniques.

\subsection{Discrete event simulation}
Discrete Event Simulation (DES) is a common simulation paradigm appreciated for its usability and good expressiveness~\cite{FUJ00}. A DES is represented by a simulated model (modelled through a set of state variables) and its evolution (modelled by a sequence of events that are processed in chronological order). Each event is timestamped (i.e.~occurs at a specific instant in the simulated time) and represents a specific change in the state variables (i.e.~a change of the simulated model state). Following this approach, the evolution of the whole simulated system is obtained through the processing of an ordered sequence of timestamped events. As an example, take a Mobile Ad-hoc NETtwork (MANET), in an Internet of Things (IoT) scenario. In this case, the events are used to model the updates of the mobile nodes positions, as well as data packets transmission among communicating nodes. Each event is tagged by a timestamp that specifies the simulated time at which it has to be processed. Otherwise, the simulation would not model correctly the evolution of the simulated system.

\subsection{Sequential DES}
In a sequential (i.e.~monolithic) simulation, a single Physical Execution Unit (PEU) is in charge of the generation of new events, managing the pending event list and processing the events that are extracted in timestamp order from the list. Hence, the simulation consists of a single executing process. While this simulator is very simple, it has limited performance. When the simulation consists of many different entities that are competing and interacting in a simulated world, a sequential program can make the task of simulating multiple and concurrent activities harder. Moreover, the scalability of the simulator is limited, both in terms of execution time (to complete the simulations runs) and complexity of the system that can be modelled~\cite{1668384-2006}.

\subsubsection{Parallel DES and PADS}
As an alternative to a sequential simulation, DES can be parallelized/distributed using a set of networked PEUs (e.g.~CPU cores, processors or hosts)~\cite{FUJ00,gda-jpdc-2017}. In this case, each PEU models only a part of the simulation model. This allows the modelling and the processing of larger and more complex simulation models with respect to DES. Each PEU manages a local pending event list. To let simulated entities interact with other simulated entities executed over other PEUs, some events are exchanged and delivered to these PEUs. This implies that a synchronization algorithm among PEUs is needed to guarantee the correct simulation execution~\cite{FUJ00,4553339-2008}. Evidence demonstrates that, in many cases, a Parallel DES (PDES) approach can speed up the simulation execution~\cite{gda-hpcs-16,gda-simpat-iot-2016,gda-jpdc-2017}.

There are many ways to implement a PDES, among them the Parallel And Distributed Simulation (PADS)~\cite{FUJ00} approach can improve execution speed, model scalability, interoperability and composability of different simulators. With respect to a sequential simulation, a PADS lacks a global model state, since each PEU manages only a part of the simulated model. According to the PADS terminology, the model components, executed on top of each PEU and interacting to advance the simulation, are called Logical Processes (LPs)~\cite{gda-jpdc-2017}.

Usually, the simulation is called ``parallel'' when the LPs are run on PEUs with a tightly-coupled configuration. Conversely, the simulation is ``distributed'' when the interconnection among the PEUs is given by a computer network (i.e.~LAN, WAN or Internet). Clearly enough, the performance of the underlying network has a significant effect on the simulator speed.  Indeed, the parallelization/distribution of the simulator does not come for free. It is important to distribute the model as evenly as possible, doing whatever is possible to minimize message passing requirements between LPs and to synchronize all the LPs to ensure a correct and consistent simulation~\cite{gda-simpat-2014}.

\subsection{Model partitioning in PADS}
In PADS, the simulated model is partitioned in a set of interacting LPs~\cite{bagrodia98}. Each LP handles a specific part of the simulation or, in other words, a set of simulated entities. This partitioning of the simulated model can be approached in many different ways. One possibility is to try to minimize the amount of network communication among LPs while doing the best to balance the workload of the LPs in the PADS execution architecture.

Over the years, many static and dynamic approaches have been evaluated to automate and enhance the partitioning of parallel and distributed simulations. We will describe them in the next section.

\section{RELATED WORK}\label{sec:related}

From the previous discussion, it should be clear that effective and scalable Internet-based simulation should embody sophisticated strategies to partition the simulation and adaptively migrate simulation entities, so that interacting simulated entities are placed within the same LPs~\cite{685264-1998}. In this section, we review the state of the art. It is worth noticing that all the approaches that are discussed in this section are not aimed at an Internet-based setup and this is the most important difference with respect to the setup investigated in this paper.

In~\cite{bagrodia98}, different partitioning schemes of the simulated region are evaluated, with the goal to assign an approximately equal number of simulated nodes to different LPs, hence reducing the amount of inter-processor messages. In this case, the partitioning schemes are static and cannot be adjusted at runtime (while the mechanism we propose is based on a dynamic approach). Another static approach for parallel conservative simulations is presented in~\cite{Boukerche:1994:SPM:182478.182586}. The authors of~\cite{847157-2000} propose a combination of static partitioning and dynamic load balancing strategies. In this case, the approach relies on a conservative synchronization algorithm and it can be used only on shared memory systems.

Many of the approaches that have been proposed to deal with load balancing are designed to address computational load balancing~\cite{gary-2-2004,gary-1-2005,Alkharboush:2013:LPH:2570454.2570889} or the communication aspects~\cite{1240666-2003,5356113-2009}, but not both of them. Moreover, very often, only optimistic synchronization (e.g.~the Time-Warp algorithm \cite{Kim1998433-1998}) is considered~\cite{4262808-2007,5356113-2009}. In many of the proposed approaches, the granularity of the load balancing mechanism is at the level of the whole LP. In other words, a whole LP is migrated from a CPU to another one. A different approach is proposed in~\cite{Schlagenhaft:1995:DLB:214282.214337} in which the LP is the container of many ``basic elements'', such as it happens in our approach. Also in this case, the synchronization relies on Time-Warp and even if it considers the communication cost in the initial partitioning, the dynamic load balancing is based only on the CPU load.

A relevant work is presented in~\cite{4262808-2007}, in which the authors propose a dynamic partitioning algorithm, which performs both computation and communication load balancing. To estimate the capacity of each host, a performance benchmark is executed before starting the simulation runs. The dynamic part of the approach is based on measurements at runtime and the migration acts on whole LPs (differently from our approach in which we migrate fine-grained model components). Another approach that takes into account both computation and communication is presented in~\cite{1348301-2004}. In this case, benchmark experiments are executed before the simulation runs and there is no adaptation at runtime, i.e.~the partitioning is static. In~\cite{Nandy:1993:PPT:158459.158465} it is assumed that there is complete knowledge of computation and communication load before the simulation is executed. 
In~\cite{Boukerche:2000:PPS:510378.510591} it is used a very specific approach that employs a two-stage parallel simulation that makes use of a conservative scheme at stage 1 and of Time-Warp at stage 2. The goal of the first stage is to collect data that will be used to partition the simulation and improve the balancing in stage 2. The improvement of the second stage is obtained by stabilizing the Time-Warp and therefore reducing the rollback overhead. In~\cite{745025-1998}, another dynamic partitioning mechanism is proposed which is based on object migration taking into account both communication and computation. A set of assumptions is made, i.e~the number of hosts running the distributed simulation is in the range from 4 to 8; the fluctuation of the network bandwidth is negligible; the migration cost of a piece of workload is proportional to the physical size of the migrating workload. Furthermore, the process of choosing which objects must be migrated is partially centralized. In should be clear that this kind of approach is not well suited for Internet-based distributed simulations.

\section{SIMULATION MODEL PARTITIONING}\label{sec:partitioning}

In this section, we introduce the partitioning scheme and the adaptive strategies we implemented to support Internet-based simulations and to improve their performance. We thus describe the Generic Adaptive Interaction Architecture (GAIA) software, how the migration has been implemented, and the heuristics we used in order to have a self-clustering scheme that is able to balance the LPs' workload.

The rationale behind the approach is that in most systems, the interactions among its parts are not randomly distributed. Rather, it is usual to observe that some communication patterns exist. These patterns are derived from the nature of the systems to be simulated and are reflected in the corresponding simulation models. When this assumption is true, it is possible to analyze such communication patterns and to find the simulation components that are involved. Through this analysis, a clustering of interacting simulated entities can be identified.

\begin{table*}[htb]
\centering
\caption{Execution testbed. All nodes equipped with Ubuntu Linux 18.04.2 LTS.\label{tab:testbed}}
\begin{tabular}{rlccll}
\hline
Label & Location & vCPU cores & GHz & RAM & Provider\\ \hline
okeanos & Greece & 4 & 2.1 & 4 GB & Okeanos Global\\
linode-SG & Singapore & 1 & 2.8 & 2 GB & Linode, LLC\\
linode-US & California (US) & 1 & 2.0 & 1 GB & Linode, LLC\\
\hline
\end{tabular}
\end{table*}

\begin{table*}[htb]
\centering
\caption{Average round-trip-response (RTT) in milliseconds between every couple of nodes in the testbed.\label{tab:latency}}
\begin{tabular}{|r|c|c|c|}
\hline
\emph{RTT} & okeanos & linode-SG & linode-US\\ \hline
okeanos     & $\circ$     & 369.7     & 218.7\\ 
linode-SG   & 369.6     & $\circ$     & 242.1\\
linode-US   & 213.0     & 369.7     & $\circ$\\
\hline
\end{tabular}
\end{table*}

\begin{table*}[htb]
\centering
\caption{Average bandwidth (Mbit/sec) between every couple of nodes in the testbed, measured using iperf.\label{tab:band}}
\begin{tabular}{|r|c|c|c|}
\hline
\emph{Mbit/sec} & okeanos & linode-SG & linode-US\\ \hline
okeanos     & $\circ$     & 17.4      & 6.47\\
linode-SG   & 35.3      & $\circ$     & 1.41\\
linode-US   & 1.88      & 1.44      & $\circ$\\
\hline
\end{tabular}
\end{table*}

\begin{figure*}[htb]
{
\centering
\includegraphics[width=0.60\textwidth]{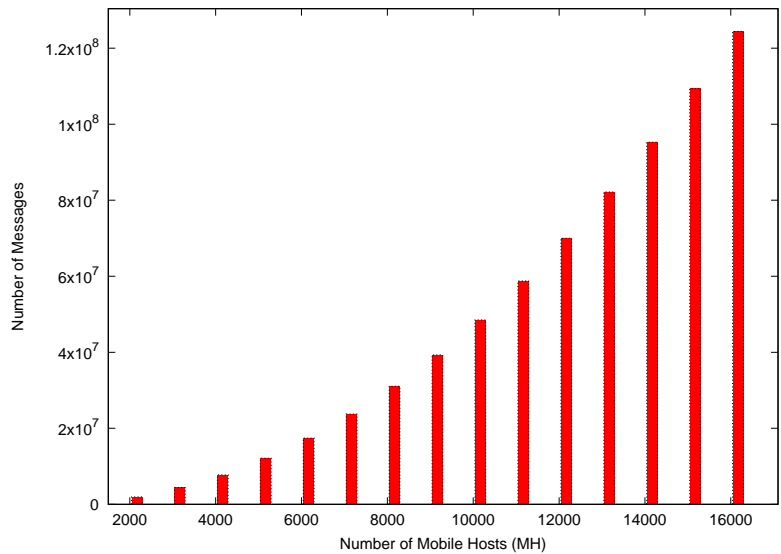}
\caption{Number of messages generated by an increasing number of simulated Mobile Hosts (MH).\label{fig:model_messages}}
}
\end{figure*}

\begin{figure*}[htb]
{
\centering
\includegraphics[width=0.60\textwidth]{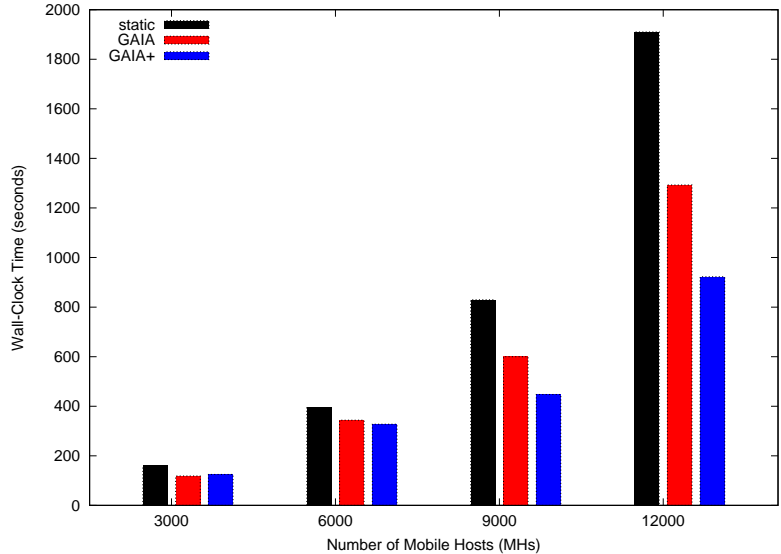}
\caption{Wall-Clock Time for completing a simulation run with an increasing number of simulated Mobile Hosts (MH). Different setups: distributed simulation with static partitioning, communication load-balancing (GAIA), communication and computational load-balancing (GAIA+). Lower is better.\label{fig:wct}}
}
\end{figure*}

\begin{figure*}[htb]
{
\centering
\includegraphics[width=0.60\textwidth]{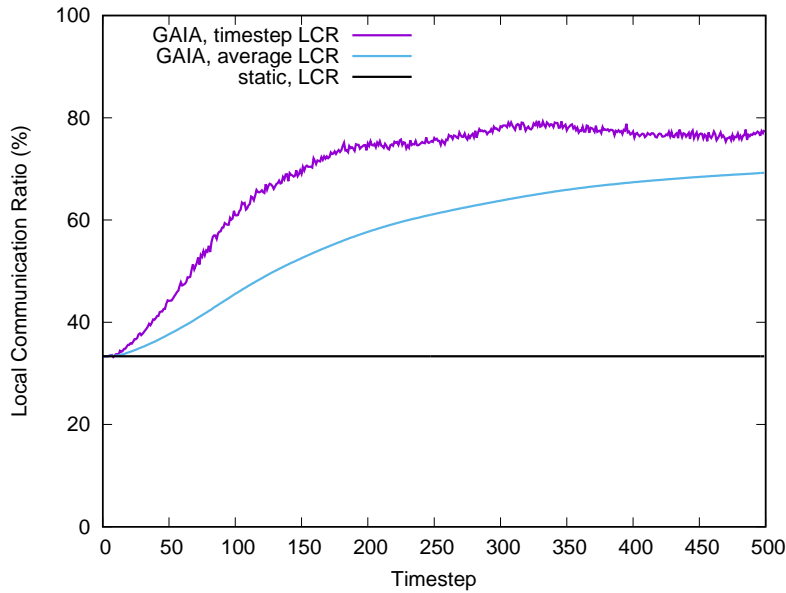}
\caption{Local Communication Ratio (LCR) and average LCR for every timestep in a simulation run. Higher is better.\label{fig:lcr}}
}
\end{figure*}

\begin{figure*}[htb]
{
\centering
\includegraphics[width=0.60\textwidth]{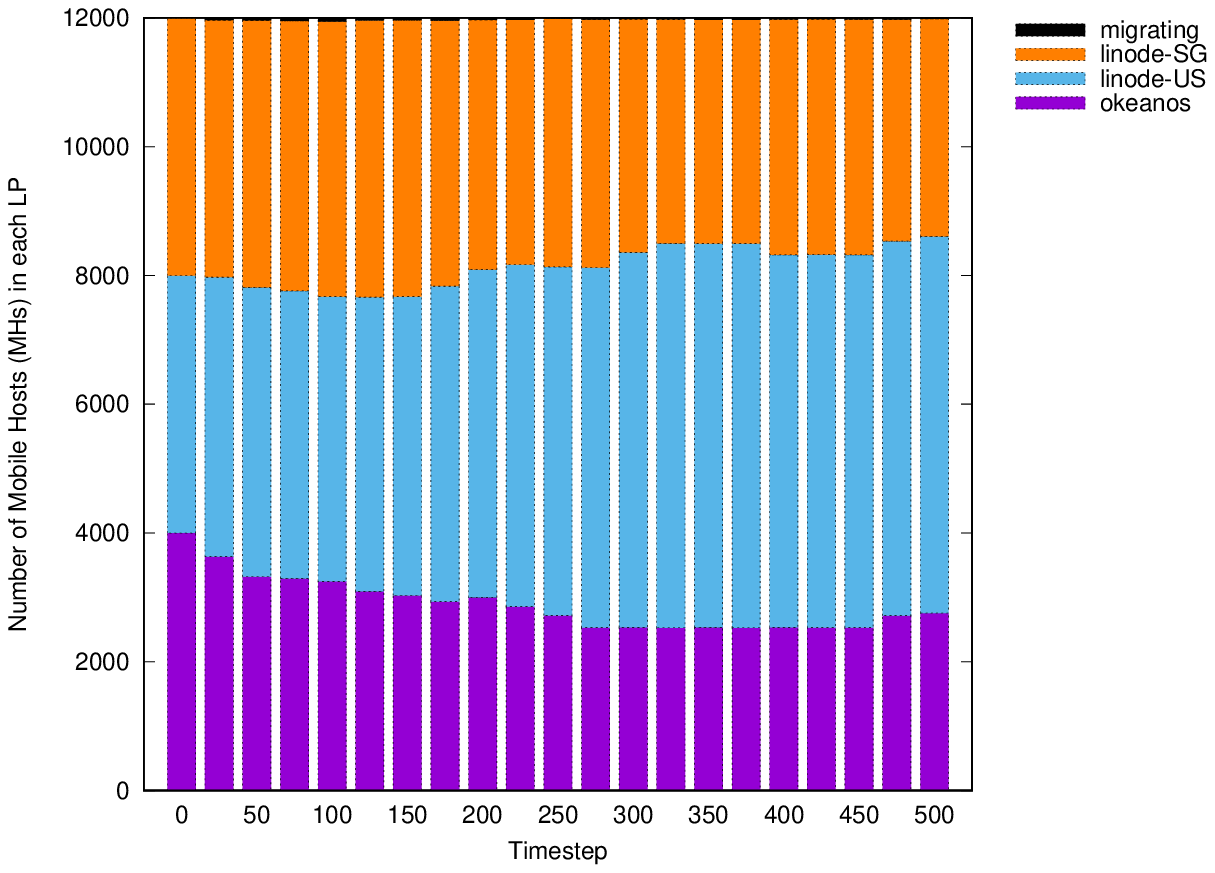}
\caption{Number of Mobile Hosts (MH) allocated on each LP during a GAIA+ simulation run.\label{fig:allocation}}
}
\end{figure*}

\subsection{GAIA basics}

The Generic Adaptive Interaction Architecture (GAIA) is a software layer built on top of the Advanced RTI System (ART\`IS) middleware~\cite{pads}. ART\`IS is a PADS middleware that provides some services such as the simulation management (i.e.~LPs coordination, simulation bootstrap and shutdown, runtime statistics), the support for the main synchronization mechanisms (e.g.~conservative and optimistic) and the communication primitives for the interaction among LPs.

On top of the middleware, GAIA provides communication APIs to let Simulated Entities (SEs) interact, and some basic utilities (e.g.~proximity detection) to speed up the model design and implementation. GAIA implements a self-clustering mechanism (i.e.~model migrations and self-clustering heuristics) that is transparent to the simulation model developer. GAIA is structured as a multi-agent system. Thus, each SE is an agent and the interactions among SEs are implemented as messages between agents. As mentioned, GAIA implements an adaptive model partitioning based on self-clustering. Its main role is to analyze the interactions among SEs and to evaluate if and where some SEs should be clustered together. Each LP is responsible for its SEs. The LP analyzes the interactions of its SEs and communicates with other LPs to identify how to cluster SEs.

\subsection{Migration implementation}
\label{Migration}
In our approach, the LP is the container of a set of SEs. The mapping between SEs and their LPs can change during the simulation run. Thus, the GAIA middleware is in charge of maintaining an updated mapping of the SEs location. Needless to say, the dynamic reallocation of SEs cannot alter, in any way, the semantic of the simulation. That is, the simulation based on adaptive partitioning must obtain the very same results as the one with static partitioning. This imposes the use of a synchronization scheme among LPs. The current implementation of GAIA is based on a time-stepped scheme in which the simulated time is divided in fixed-size timesteps and each LP can proceed to the next timestep only when all other LPs have completed the elaboration of the current timestep~\cite{FUJ00}. The decision of migrating some SEs is based on a self-clustering heuristic function that is implemented in each LP (some details will be given in the next subsection).  Once a SE migration is triggered, all the LPs are informed, in order to maintain a consistent view of the SEs placement. A critical point is that during a SE migration, all messages that have to be delivered to that SE have to be routed to the final LP, where the SE will be hosted. This imposes some constraints and control checks on the synchronization algorithm. Implementation details about this scheme can be found in \cite{gda-simpat-2017}.

\subsection{Self-clustering heuristics}
\label{Heuristics}
The goal of the self-clustering heuristic is to identify groups of SEs, and group these SEs into LPs, so as to minimize the amount of inter-group (i.e.~inter-LP) communications (and the number of migrations). A main constraint of the heuristic function is to rely on local information only, and avoid every kind of centralization. Moreover, the whole process should require as little computation as possible, for better performance and scalability. This last point is very crucial, given that many simulated systems are often modeled by a huge number of SEs and each of them has to be analyzed by the clustering heuristic. Thus, the self-clustering heuristic scheme is based on the analysis of the communication pattern of SEs (e.g.~amount of intra-LP vs.~inter-LP communication). More in detail, if most of the interactions of a given SE are delivered outside the local LP then this SE is a candidate for being migrated by the simulation middleware. The real migration will be triggered only if the cost of migrating the SE's data structures is expected to be balanced by the communication cost reduction given by the migration.

The partitioning of the simulated model involves both communication and computational aspects. In fact, if only communication is considered, then the best strategy would be to cluster all the SEs in the same LP, i.e.~monolithic simulation. GAIA implements more load balancing aware clustering schemes, and it can be configured, based on the specific needs.

In particular, a symmetric load balancing scheme tries to avoid imbalance caused by SEs migrations. Assuming that the simulation of different SEs have similar computational requirements, the workload of each LP is proportional to the amount of SEs it manages. Therefore, symmetric load balancing tries to assign a fair amount of SEs to all LPs, and migrations, that would cause imbalances of the LPs loads, are not executed. 

In the asymmetric load balancing scheme, in a LP the inbound and outbound migrations may be imbalanced. The scheme hence deals also with heterogeneous distributed systems, where PEUs may have different execution speeds, different background communication/computational loads, etc. This scheme (referred to as GAIA+) adapts the model partitioning to the runtime characteristics of the execution architecture. Live data are collected during the simulation execution instrumenting the synchronization mechanism. This data is then used to migrate SEs from slower LPs (reducing their load), to faster ones (that will slow down). In other words, the time-stepped synchronization mechanisms is used to coordinate the execution of the different LPs (and PEUs) that participate in the distributed simulation to avoid causality errors during the simulation execution. Moreover, it can be used to evaluate the execution speed of the different execution components. This fits very well with the migration mechanism that can be tuned to block migrations from the fast parts of the simulation and to trigger many of them from the slow PEUs up to when the desired balancing is obtained. The reader that is interested in the inner working of GAIA and GAIA+ can refer to \cite{gda-simpat-2014} and \cite{gda-simpat-2017} for a more detailed description.

\section{PERFORMANCE EVALUATION}\label{sec:perf}

In this section, a simple wireless model is evaluated on a real Internet-based distributed simulation setup. The GAIA/ART\`IS simulation middleware and the wireless model used for building the adaptive distributed simulator are freely available from the research group Web site~\cite{pads}.

\subsection{Simulation model}
We consider a time-stepped, Mobile Ad-hoc wireless Network (MANET) model of $3000$-$12000$ Mobile Hosts (MHs). In the simulated model each MH is implemented as a SE. The simulated scenario is a two-dimensional area ($10000 \times 10000$ units) with periodic boundary conditions, in which each MH follows a random waypoint mobility model ($\mathit{max speed}=5$ space-units/time-unit, $100\%$ of nodes move at each timestep). The communication model is very simple and does not consider the details of low level medium access control. At each timestep, a random subset of $20\%$ of the MHs broadcasts a ping message to all nodes that are within the transmission radius of $250$ space units. Each simulation run is composed of $500$ timesteps and all the results reported in this performance evaluation are averages of multiple-independent runs. This model captures both the dynamicity of the simulated systems and the space locality aspects of wireless communication. At the simulation bootstrap, the MHs have been randomly partitioned in $3$ PEUs, each one running a single LP. As shown in Table~\ref{tab:testbed} the PEU are running on virtual machines executed in different parts of the globe.

It is clear that this MANET simulation model is a (near to) worst case scenario for the clustering and migration management scheme employed in the distributed simulation. In fact, the continuous movement of simulated MHs, during the whole simulation, forces the adaptive migration strategy to continuously reconfigure and migrate simulated entities among physical LPs. Thus, obtained results can be considered as a lower bound of the performance that can be obtained simulating models that are more static.

Figure~\ref{fig:model_messages} shows the (average) number of messages, generated by our simulation model, when varying the number of simulated MHs. It is possible to observe a more than linear trend in the amount of messages, resulting in a higher complexity of the model, when augmenting the amount of SEs. It is thus interesting to observe if the Internet-based simulator scales with respect to this parameter.

\subsection{Simulation setup}\label{sec:res}

Tables~\ref{tab:testbed}--\ref{tab:band} describe the simulation setup and the related network parameters. In particular, as shown in Table~\ref{tab:testbed}, we used three hosts widely distributed around the globe, i.e.~Greece, Singapore, and the US. The hosts have a different number of available vCPUs and RAM, as reported in the table.

Table~\ref{tab:latency} shows the average round-trip-response (RTT) between every couple of nodes in the testbed and Table~\ref{tab:band} reports the average bandwidth (Mbit/sec) between every couple of nodes. It can be noticed that we are in the presence of a diverse and asymmetric communication among the hosts used in the testbed. We assume that this might have an impact on the performance.

\subsection{Results}

In Figure~\ref{fig:wct}, we report the Wall-Clock Time (WCT) to complete a simulation run in the distributed execution testbed described above and in the presence of an increasing number of MHs. As expected, the time required for a simulation run in the presence of a \emph{static} allocation of the MHs on the LPs (i.e.~black bars in the figure) follows the number of delivered messages in the simulation model (see Figure~\ref{fig:model_messages}). This limits the scalability of the simulator. A reduction of the WCT can be obtained using a mechanism that clusters the communications (i.e.~GAIA, red bars). The best results are obtained when exploiting the adaptive reallocation scheme that considers both communication clustering and computational load-balancing (i.e.~GAIA+, blue bars). When the number of MHs is low (i.e.~$3000$), the gain in terms of WCT obtained by GAIA is $26.07$\% while GAIA+ is unable to provide a further speedup ($22.11$\%). This happens because the computational load of $3000$ MHs is so low that the performance of the distributed simulation run are mostly dominated by the communication cost. In other words, GAIA+ is unable to perform an effective computational load-balancing since there is no relevant computation to balance. This does not happen when the number of simulated MHs is increased. In fact, with $12000$ MHs the speedup provided by both GAIA and GAIA+ is evident. In this case, GAIA is able to reduce the WCT of $32.31$\%, while GAIA+ provides a gain of $51.71$\%.\\

As described above, GAIA works by dynamically clustering the groups of interacting MHs in the same LP. This can be easily done in simulation models in which there is a high level of communication locality, but it is harder in models in which the simulated entities do not have group behaviors. A very simple way to evaluate the efficiency of the communication clustering heuristic is to consider how much it is able to increase the amount of local (intra-LP) communication, with respect to the remote (inter-LP) communication. Figure~\ref{fig:lcr} shows the Local Communication Ratio (LCR) (i.e.~percentage of local communication with respect to the total amount of communications) provided by GAIA with respect to the \emph{static} distributed simulation (black line). Even if the simulation model used in this performance evaluation is quite dynamic, GAIA provides a good LCR (a little below $75$\%), with a sharp increase in the first simulation timesteps. Obviously, in terms of average LCR, the longer the simulation run, the higher the average LCR.\\

To investigate the behavior of GAIA+, we need to consider the number of MHs that is allocated in every LP during the simulation. As shown in Figure~\ref{fig:allocation}, at the simulation bootstrap every LP allocates the same number of MHs (in this case, $4000$). The goal of GAIA+ is to provide both communication clustering and computational load-balancing. This means that if in the distributed execution testbed there is one LP that slows down the simulation (i.e.~it is overloaded or with a laggy network), then some MHs must be migrated to other LPs. In most cases, this removes the bottleneck and speeds up the execution. As reported in the figure, during the simulation run it happens that both \emph {okeanos} and \emph{linode-SG} reduce the MHs moving some of them to \emph{linode-US}.

It is worth noticing that this result is partially unexpected if we consider the specifications of the virtual machines that have been used (Table~\ref{tab:testbed}). In other words, we were expecting that \emph{linode-US} would have migrated a lot of MHs outbound to compensate the low specifications of this virtual instance. In practice, this does not happen since GAIA+ considers the execution speed of each LP when compared with the rest of the execution architecture. For this reason, even if \emph{linode-US} is slow in terms of computation, it can be perceived as faster than other LPs, due to the characteristics of the communication network.

\section{CONCLUDING REMARKS}\label{sec:conc}

In this paper, we focused on Internet-based distributed simulation, an important yet neglected problem of PADS. We described the related issues and presented a distributed simulator we implemented. The simulator embodies an adaptive partition scheme, with the aim of reducing both computation and communication workload, thus gaining in scalability and performance. In particular, the employed heuristics aim at clustering interacting SEs with the idea of properly balancing the workload at LPs, while still minimizing the inter-LP communication.

To the best of our knowledge, this is the first study that performed this kind of evaluation on an Internet-based execution testbed. We placed LPs in three continents. Thus, the network latencies played an important role on the performance of the simulation. The results showed that the self-clustering mechanism provided by GAIA was able to speed up the simulation runs and that, in most cases, the computational load-balancing provided by GAIA+ provided further benefits in terms of simulation execution speed.

As a future work, we plan to perform a more complex and wide experimental campaign on Internet-based simulation. It would be interesting to analyze the performance of the distributed simulator with a higher number of PEU, varying their heterogeneity. It would also be interesting to consider different simulation models. Indeed, the simulation model we employed in this study was a sort of worst case scenario, with SEs that continuously change their interactions with others. This imposes a continuous reconfiguration of the LPs, thus triggering the migration of certain SEs.

\small{
\bibliographystyle{abbrv}
\bibliography{biblio}  
}

\end{document}